\documentclass[aps,prl,twocolumn,superscriptaddress,floatfix,nofootinbib,showpacs,amsmath,amssymb]{revtex4-1}
\usepackage[colorlinks=true,pdfstartview=FitV,linkcolor=blue,citecolor=blue,urlcolor=blue]{hyperref}
\usepackage{wasysym}
\usepackage[separate-uncertainty,retain-explicit-plus,per-mode=symbol,binary-units]{siunitx}
\usepackage{array,mathtools,amssymb,dcolumn}
\usepackage[below]{placeins}
\usepackage[dvipsnames, table]{xcolor}
\usepackage{tikz}
\usepackage{multirow}
\usepackage{afterpage}
\usepackage{lineno}
\usepackage{paralist}
\usepackage{listings}
\usepackage{array}
\usepackage{cancel}
\usepackage{tabularx}
\newcommand\T{\rule{0pt}{2.6ex}}       
\newcommand\B{\rule[-1.2ex]{0pt}{0pt}} 
\usepackage[version=4]{mhchem}
\usepackage{calc}
\newcommand{\beq}{\begin{equation}}
\newcommand{\eeq}{\end{equation}}
\newcommand{\bea}{\begin{eqnarray}}
\newcommand{\eea}{\end{eqnarray}}

\usepackage[separate-uncertainty,retain-explicit-plus,per-mode=symbol,binary-units]{siunitx}
\usepackage{calc}
\newcommand{\refcite}[1]{\mbox{Ref.~\cite{#1}}}
\newcommand{\refscite}[1]{\mbox{Refs.~\cite{#1}}}
\newcommand{\refsec}[1]{\mbox{Sec.~\ref{#1}}}
\newcommand{\reftab}[1]{\mbox{Tab.~\ref{#1}}}
\newcommand{\reffig}[1]{\mbox{Fig.~\ref{#1}}}
\newcommand{\refeqn}[1]{\mbox{Eq.~\ref{#1}}}

\newcommand{\Geant}{\texttt{Geant4}}

\newcommand{\RAT}{\texttt{RAT}}

\DeclareSIUnit\c{\mbox{$c$}}
\DeclareSIUnit\week{w}
\DeclareSIUnit\year{yr}
\DeclareSIUnit\yr{yr}
\DeclareSIUnit\yr{yr}
\DeclareSIUnit\standard{std}
\DeclareSIUnit\str{sr}
\DeclareSIUnit\ppm{ppm}
\DeclareSIUnit\ppb{ppb}
\DeclareSIUnit\ppt{ppt}
\DeclareSIUnit\pe{PE}
\DeclareSIUnit\spe{SPE}
\DeclareSIUnit\ev{events}
\DeclareSIUnit\hit{hit}
\DeclareSIUnit\hits{hits}
\DeclareSIUnit\bin{(\mbox{5-PE}~bin)}
\DeclareSIUnit\sgm{\mbox{$\sigma$}}
\DeclareSIUnit\rms{RMS}
\DeclareSIUnit\keVr{\mbox{keV$_{\rm nr}$}}
\DeclareSIUnit\keVee{\mbox{keV$_{\rm ee}$}}
\DeclareSIUnit\MeVee{\mbox{MeV$_{\rm ee}$}}
\DeclareSIUnit\ph{photons}
\DeclareSIUnit\pm{PMT}
\DeclareSIUnit\inch{''}
\DeclareSIUnit\bit{bit}
\DeclareSIUnit\sample{S}
\DeclareSIUnit\barn{b}
\DeclareSIUnit\bara{bar}
\DeclareSIUnit\Curie{Ci}
\DeclareSIUnit\psi{psi}
\DeclareSIUnit\ms{\milli\second}
\DeclareSIUnit\mK{\milli\kelvin}
\DeclareSIUnit\micron{\micro\metre}
\DeclareSIUnit\liveday{\mbox{live-days}}
\DeclareSIUnit\tonneday{\mbox{tonne$\cdot$day}}
\DeclareSIUnit\days{\mbox{days}}

\usepackage[symbol*]{footmisc}
\DefineFNsymbolsTM{otherfnsymbols}{
  \textdagger    \dagger
  \textasteriskcentered *
  \textbardbl    \|
  \textparagraph \mathparagraph
  \textbullet \circ
  \textdaggerdbl \ddagger
}
\setfnsymbol{otherfnsymbols}

\newcommand{\SnolabDepth}{\SI{2}{\km}}

\newcommand{\AVPMTsNum}{255}

\newcommand{\larmasserror}{\SI{3279 \pm 96}{\kg}}

\newcommand{\PaperThreeAVSurfaceNEventUpperLimit}{\num{2.3}}

\newcommand{\ArThreeNineTotalActivity}{\SI{3.3\pm0.3}{\kilo\becquerel}}

\newcommand{\isotope}[2]{\mbox{$^{#2}$#1}}
\newcommand{\SNOLAB}{\mbox{SNOLAB}}
\newcommand{\DEAP}{\mbox{DEAP-3600}}

\newcommand{\AV}{\mbox{AV}}
\newcommand{\MV}{\mbox{MV}}

\newcommand{\SNOp}{\mbox{SNO+}}
\newcommand{\JUNO}{\mbox{JUNO}}
\newcommand{\MATHUSLA}{\mbox{MATHUSLA}}

\newcommand{\PMTs}{\mbox{PMTs}}

\newcommand{\WIMP}{\mbox{WIMP}}

\newcommand{\WIMPs}{\mbox{WIMPs}}
\newcommand{\DM}{\mbox{DM}}

\newcommand{\LAr}{\mbox{LAr}}

\newcommand{\ROI}{\mbox{ROI}}
\newcommand{\ROIs}{\mbox{ROIs}}

\newcommand{\eg}{\mbox{\emph{e.g.}}}
\newcommand{\CL}{C.~L.}

\newcommand{\RhoDMSymbol}{\mbox{$\rho_{\chi}$}}
\newcommand{\RhoDMValue}{\SI{0.3}{\GeV\per\square\c\per\cubic\cm}}

\newcommand{\WIMPMassSymbol}{\mbox{m$_\chi$}}

\newcommand{\NinetyPerCentCL}{\mbox{\SI{90}{\percent}~\CL}}

\newcommand{\alpds}{\mbox{$\alpha$-decays}}

\newcommand{\betds}{\mbox{$\beta$-decays}}

\newcommand{\ngamma}{\mbox{($n,\gamma$)}}

\newcommand{\subeventN}{\mbox{$\text{N}_\text{peaks}$}}
\newcommand{\FPrompt}{\mbox{F$_{\text{prompt}}$}}

\newcommand{\PE}{\mbox{PE}}
\newcommand{\PEs}{\mbox{PEs}}

\newcommand{\larmassapprox}{\SI{3.3}{\tonne}}

\newcommand{\QPEWindow}{\SI{10}{\us}}

\newcommand{\PrescaleFactor}{\SI{99}{\percent}}

\newcommand{\PrescaleRangeEne}{\SIrange{50}{565}{\keVee}}

\newcommand{\TimeFitWindowEndNum}{150}
\newcommand{\TimeFitWindowEnd}{\SI{\TimeFitWindowEndNum}{\ns}}

\newcommand{\PoTwelveHalfLife}{\SI{300}{\ns}}

\newcommand{\MIMPGenerationRadius}{\SI{1.5}{\m}}
\newcommand{\MIMPAtmRadius}{\SI{80}{\km}}
\newcommand{\MIMPAtmNtwoConc}{\SI{79}{\percent}}
\newcommand{\MIMPAtmOtwoConc}{\SI{21}{\percent}}

\newcommand{\MIMPROIOnePERange}{\numrange{4000}{20000}}
\newcommand{\MIMPROITwoPERange}{\numrange{20000}{30000}}
\newcommand{\MIMPROIThreePERange}{\numrange{30000}{70000}}
\newcommand{\MIMPROIFourPERange}{\numrange{70000}{4e8}}

\newcommand{\MIMPROIOneEneRange}{\numrange{0.5}{2.9}}
\newcommand{\MIMPROITwoEneRange}{\numrange{2.9}{4.4}}
\newcommand{\MIMPROIThreeEneRange}{\numrange{4.4}{10.4}}
\newcommand{\MIMPROIFourEneRange}{\numrange{10.4}{60000}}
\newcommand{\MIMPROIOneSubEventNMin}{\num{7}} 
\newcommand{\MIMPROITwoSubEventNMin}{\num{5}} 
\newcommand{\MIMPROIThreeSubEventNMin}{\num{4}}
\newcommand{\MIMPROIFourSubEventNMin}{\num{0}} 
\newcommand{\MIMPROIOneFpMax}{\num{0.10}} 
\newcommand{\MIMPROITwoFpMax}{\num{0.10}} 
\newcommand{\MIMPROIThreeFpMax}{\num{0.10}}
\newcommand{\MIMPROIFourFpMax}{\num{0.05}} 
\newcommand{\MIMPROIOneNbkgd}{\num{4\pm3e-2}} 
\newcommand{\MIMPROITwoNbkgd}{\num{6\pm1e-4}} 
\newcommand{\MIMPROIThreeNbkgd}{\num{6\pm2 e-4}} 
\newcommand{\MIMPROIFourNbkgd}{\num{10\pm3e-3}} 
\newcommand{\MIMPROIOneNobs}{\num{0}} 
\newcommand{\MIMPROITwoNobs}{\num{0}} 
\newcommand{\MIMPROIThreeNobs}{\num{0}} 
\newcommand{\MIMPROIFourNobs}{\num{0}} 
\newcommand{\AmBeSourceActivity}{\SI{4.6\pm0.7}{\kilo\hertz}} 
\newcommand{\MIMPAmBeLivetime}{\SI{3.8}{\hour}}
 
\newcommand{\MIMPSubeventNMaxDisagreement}{\SI{5}{\percent}}
\newcommand{\MIMPLiveTimeNoErrors}{\SI{813}{\day}}

\newcommand{\MIMPLiveTimeNoUnits}{{\num{813 \pm 8}}}
\newcommand{\MIMPOpenDatasetLiveTime}{\SI{9}{\day}} 
\newcommand{\MIMPMuonCutWindow}{\mbox{\SIrange[range-phrase={,}\ , range-units = brackets, open-bracket = [,close-bracket = ]]{-10}{90}{\micro\second}}} 
\newcommand{\MIMPMuonSidebandLiveTime}{\SI{6}{\day}} 
\newcommand{\MIMPAcceptanceCFFTR}{\SI{99.1\pm0.1}{\percent}} 
\newcommand{\MIMPAcceptanceFmaxpe}{\SI{86.5\pm0.3}{\percent}}
\newcommand{\MIMPCFTTRCut}{\SI{5}{\percent}}
\newcommand{\MIMPFmaxpeCut}{\SI{5}{\percent}}
\newcommand{\MIMPLYCorrFactor}{\num{0.9\pm0.1}}
\newcommand{\PreviousDMMassLimit}{\SI{6e17}{\GeV\per\square\c}}
\newcommand{\MIMPMinimumFpAcceptanceROIFour}{\SI{35}{\percent}}
\newcommand{\MIMPSearchStartDate}{November 4, 2016}
\newcommand{\MIMPSearchEndDate}{March 8, 2020}
\newcommand{\MIMPTotalBkgdExpectation}{\num{0.05\pm0.03}}
\newcommand{\NumberExpectedSignalSymbol}{\mbox{$\mu_s$}}
\newcommand{\NumberExpectedBackgroundSymbol}{\mbox{$\mu_b$}}

\def\Ndark{{\rm N}_{\rm D}}
\def\rdark{{\rm r}_{\rm D}}
\def\Rdark{{\rm R}_{\rm D}}
\def\mdark{\mbox{m$_{\rm D}$}}

\def\mT{\text{m}_{\rm T}}

\def\sigmaNX{\sigma_{\rm n \chi}}
\def\sigmaTX{\sigma_{\rm T \chi}}

\allowdisplaybreaks

\begin{document}
\setcounter{secnumdepth}{5}
\title{First direct detection constraints on Planck-scale mass dark matter with multiple-scatter signatures using the DEAP-3600 detector}
\newcommand{\UofA}{Department of Physics, University of Alberta, Edmonton, Alberta, T6G 2R3, Canada}
\newcommand{\AsC}{AstroCeNT, Nicolaus Copernicus Astronomical Center, Polish Academy of Sciences, Rektorska 4, 00-614 Warsaw, Poland}
\newcommand{\CNL}{Canadian Nuclear Laboratories, Chalk River, Ontario, K0J 1J0, Canada}
\newcommand{\CIEMAT}{Centro de Investigaciones Energ\'eticas, Medioambientales y Tecnol\'ogicas, Madrid 28040, Spain}
\newcommand{\CU}{Department of Physics, Carleton University, Ottawa, Ontario, K1S 5B6, Canada}
\newcommand{\LNGSA}{INFN Laboratori Nazionali del Gran Sasso, Assergi (AQ) 67100, Italy}
\newcommand{\RHUL}{Royal Holloway University London, Egham Hill, Egham, Surrey TW20 0EX, United Kingdom}
\newcommand{\LU}{Department of Physics and Astronomy, Laurentian University, Sudbury, Ontario, P3E 2C6, Canada}
\newcommand{\UNAM}{Instituto de F\'isica, Universidad Nacional Aut\'onoma de M\'exico, A.\,P.~20-364, M\'exico D.\,F.~01000, M\'exico}
\newcommand{\INFN}{INFN Napoli, Napoli 80126, Italy}
\newcommand{\PRISMA}{PRISMA$^+$, Cluster of Excellence and Institut f\"ur Kernphysik, Johannes Gutenberg-Universit\"at Mainz, 55128 Mainz, Germany}
\newcommand{\PU}{Physics Department, Princeton University, Princeton, NJ 08544, USA}
\newcommand{\QU}{Department of Physics, Engineering Physics, and Astronomy, Queen's University, Kingston, Ontario, K7L 3N6, Canada}
\newcommand{\RAL}{Rutherford Appleton Laboratory, Harwell Oxford, Didcot OX11 0QX, United Kingdom}
\newcommand{\SL}{SNOLAB, Lively, Ontario, P3Y 1N2, Canada}
\newcommand{\Sussex}{University of Sussex, Sussex House, Brighton, East Sussex BN1 9RH, United Kingdom}
\newcommand{\TRIUMF}{TRIUMF, Vancouver, British Columbia, V6T 2A3, Canada}
\newcommand{\TUM}{Department of Physics, Technische Universit\"at M\"unchen, 80333 Munich, Germany}
\newcommand{\Napoli}{Physics Department, Universit\`a degli Studi ``Federico II'' di Napoli, Napoli 80126, Italy}
\newcommand{\INAF}{Astronomical Observatory of Capodimonte, Salita Moiariello 16, I-80131 Napoli, Italy}
\newcommand{\LBLNSD}{Currently: Nuclear Science Division, Lawrence Berkeley National Laboratory, Berkeley, CA 94720}
\newcommand{\kurchatov}{National Research Centre Kurchatov Institute, Moscow 123182, Russia}
\newcommand{\MEPhI}{National Research Nuclear University MEPhI, Moscow 115409, Russia}
\newcommand{\MI}{Arthur B. McDonald Canadian  Astroparticle Physics Research Institute, Queen's University, Kingston ON K7L 3N6,Canada}
\newcommand{\PI}{Perimeter Institute for Theoretical Physics, Waterloo ON N2L 2Y5, Canada}
\newcommand{\UdSCag}{Physics Department, Universit\`a degli Studi di Cagliari, Cagliari  09042, Italy}
\newcommand{\INFNCag}{INFN Cagliari, Cagliari 09042, Italy}
\newcommand{\NCNR}{BP2, National Centre for Nuclear Research, ul. Pasteura 7, 02-093 Warsaw, Poland}
\newcommand{\Nikhef}{Currently: Nikhef and the University of Amsterdam, Science Park, 1098XG Amsterdam, Netherland}

\affiliation{\UofA}
\affiliation{\AsC}
\affiliation{\UdSCag}
\affiliation{\CNL}
\affiliation{\CU}
\affiliation{\CIEMAT}
\affiliation{\Napoli}
\affiliation{\INAF}
\affiliation{\INFNCag}
\affiliation{\LNGSA}
\affiliation{\INFN}
\affiliation{\LU}
\affiliation{\UNAM}
\affiliation{\NCNR}
\affiliation{\kurchatov}
\affiliation{\MEPhI}
\affiliation{\PU}
\affiliation{\PRISMA}
\affiliation{\QU}
\affiliation{\RHUL}
\affiliation{\SL}
\affiliation{\Sussex}
\affiliation{\TRIUMF}
\affiliation{\TUM}
\affiliation{\MI}

\author{P.~Adhikari}\affiliation{\CU} 
\author{R.~Ajaj}\affiliation{\CU}\affiliation{\MI}
\author{M.~Alp\'izar-Venegas}\affiliation{\UNAM}
\author{D.\,J.~Auty}\affiliation{\UofA}
\author{H.~Benmansour}\affiliation{\QU}  
\author{C.\,E.~Bina}\affiliation{\UofA}\affiliation{\MI}
\author{W.~Bonivento}\affiliation{\INFNCag}
\author{M.\,G.~Boulay}\affiliation{\CU}
\author{M.~Cadeddu}\affiliation{\UdSCag}\affiliation{\INFNCag}
\author{B.~Cai}\affiliation{\CU}\affiliation{\MI} 
\author{M.~C\'ardenas-Montes}\affiliation{\CIEMAT}
\author{S.~Cavuoti}\affiliation{\INAF}\affiliation{\Napoli}\affiliation{\INFN}
\author{Y.~Chen}\affiliation{\UofA}
\author{B.\,T.~Cleveland}\affiliation{\SL}\affiliation{\LU}
\author{J.\,M.~Corning}\affiliation{\QU} 
\author{S.~Daugherty}\affiliation{\LU} 
\author{P.~DelGobbo}\affiliation{\CU}\affiliation{\MI} 
\author{P.~Di~Stefano}\affiliation{\QU} 
\author{L.~Doria}\affiliation{\PRISMA} 
\author{M.~Dunford}\affiliation{\CU}
\author{E.~Ellingwood}\affiliation{\QU}
\author{A.~Erlandson}\affiliation{\CU}\affiliation{\CNL}
\author{S.\,S.~Farahani}\affiliation{\UofA} 
\author{N.~Fatemighomi}\affiliation{\SL}\affiliation{\RHUL}
\author{G.~Fiorillo}\affiliation{\Napoli}\affiliation{\INFN}
\author{D.~Gallacher}\affiliation{\CU}
\author{P.~Garc\'ia~Abia}\affiliation{\CIEMAT}
\author{S.~Garg}\affiliation{\CU}
\author{P.~Giampa}\affiliation{\TRIUMF}
\author{D.~Goeldi}\affiliation{\CU}\affiliation{\MI}
\author{P.~Gorel}\affiliation{\SL}\affiliation{\LU}\affiliation{\MI}
\author{K.~Graham}\affiliation{\CU}
\author{A.~Grobov}\affiliation{\kurchatov}\affiliation{\MEPhI}
\author{A.\,L.~Hallin}\affiliation{\UofA}
\author{M.~Hamstra}\affiliation{\CU}
\author{T.~Hugues}\affiliation{\AsC} 
\author{A.~Ilyasov}\affiliation{\kurchatov}\affiliation{\MEPhI} 
\author{A.~Joy}\affiliation{\UofA}\affiliation{\MI}
\author{B.~Jigmeddorj}\affiliation{\CNL}
\author{C.\,J.~Jillings}\affiliation{\SL}\affiliation{\LU}
\author{O.~Kamaev}\affiliation{\CNL}
\author{G.~Kaur}\affiliation{\CU}
\author{A.~Kemp}\affiliation{\QU}\affiliation{\RHUL}
\author{I.~Kochanek}\affiliation{\LNGSA}
\author{M.~Ku{\'z}niak}\affiliation{\AsC}\affiliation{\CU}\affiliation{\MI}
\author{M.~Lai}\affiliation{\UdSCag}\affiliation{\INFNCag}
\author{S.~Langrock}\affiliation{\LU}\affiliation{\MI}
\author{B.~Lehnert}\altaffiliation{\LBLNSD}\affiliation{\CU}
\author{A. Leonhardt}\affiliation{\TUM}
\author{N.~Levashko}\affiliation{\kurchatov}\affiliation{\MEPhI} 
\author{X.~Li}\affiliation{\PU}
\author{M.~Lissia}\affiliation{\INFNCag}
\author{O.~Litvinov}\affiliation{\TRIUMF}
\author{J.~Lock}\affiliation{\CU}
\author{G.~Longo}\affiliation{\Napoli}\affiliation{\INFN}
\author{I.~Machulin}\affiliation{\kurchatov}\affiliation{\MEPhI} 
\author{A.\,B.~McDonald}\affiliation{\QU}
\author{T.~McElroy}\affiliation{\UofA}
\author{J.\,B.~McLaughlin}\affiliation{\RHUL}\affiliation{\TRIUMF}
\author{C.~Mielnichuk}\affiliation{\UofA}
\author{L.~Mirasola}\affiliation{\UdSCag}
\author{J.~Monroe}\affiliation{\RHUL}
\author{G.~Olivi\'ero}\affiliation{\CU}\affiliation{\MI} 
\author{S.~Pal}\affiliation{\UofA}\affiliation{\MI} 
\author{S.\,J.\,M.~Peeters}\affiliation{\Sussex}
\author{M.~Perry}\affiliation{\CU}
\author{V.~Pesudo}\affiliation{\CIEMAT}
\author{E.~Picciau}\affiliation{\INFNCag}\affiliation{\UdSCag}
\author{M.-C.~Piro}\affiliation{\UofA}\affiliation{\MI}
\author{T.\,R.~Pollmann}\altaffiliation{\Nikhef}\affiliation{\TUM}
\author{N.~Raj}\affiliation{\TRIUMF}
\author{E.\,T.~Rand}\affiliation{\CNL}
\author{C.~Rethmeier}\affiliation{\CU}
\author{F.~Reti\`ere}\affiliation{\TRIUMF}
\author{I. Rodr\'iguez-Garc\'ia}\affiliation{\CIEMAT}
\author{L.~Roszkowski}\affiliation{\AsC}\affiliation{\NCNR}
\author{J.\,B.~Ruhland}\affiliation{\TUM}
\author{E.~Sanchez~Garc\'ia}\affiliation{\CIEMAT}
\author{T.~S\'anchez-Pastor}\affiliation{\CIEMAT}
\author{R.~Santorelli}\affiliation{\CIEMAT}
\author{S.~Seth}\affiliation{\CU}\affiliation{\MI}
\author{D.~Sinclair}\affiliation{\CU}
\author{P.~Skensved}\affiliation{\QU}
\author{B.~Smith}\affiliation{\TRIUMF}
\author{N.\,J.\,T.~Smith}\affiliation{\SL}\affiliation{\LU}
\author{T.~Sonley}\affiliation{\SL}\affiliation{\MI}
\author{R.~Stainforth}\affiliation{\CU}
\author{M.~Stringer}\affiliation{\QU}\affiliation{\MI} 
\author{B.~Sur}\affiliation{\CNL}
\author{E.~V\'azquez-J\'auregui}\affiliation{\UNAM}\affiliation{\LU}
\author{S.~Viel}\affiliation{\CU}\affiliation{\MI}
\author{J.~Walding}\affiliation{\RHUL}
\author{M.~Waqar}\affiliation{\CU}\affiliation{\MI}
\author{M.~Ward}\affiliation{\QU}\affiliation{\SL}
\author{S.~Westerdale}\affiliation{\INFNCag}\affiliation{\CU}
\author{J.~Willis}\affiliation{\UofA}
\author{A.~Zu\~niga-Reyes}\affiliation{\UNAM}
\collaboration{DEAP Collaboration}\email{deap-papers@snolab.ca}\noaffiliation

\begin{abstract}
Dark matter with Planck-scale mass ($\simeq\SI{e19}{\GeV\per\square\c}$) arises in well-motivated theories and could be produced by several cosmological mechanisms.
A search for multi-scatter signals from supermassive dark matter was performed with a blind analysis of data collected over a \MIMPLiveTimeNoErrors\ live time with \DEAP, a \larmassapprox\ single-phase liquid argon-based detector at \SNOLAB.
No candidate signals were observed, leading to the first direct detection constraints on Planck-scale mass dark matter. Leading limits 
constrain dark matter masses between \SIrange[range-phrase=\mbox{\ and\ }]{8.3e6}{1.2e19}{\GeV\per\square\c}, and \isotope{Ar}{40}-scattering cross sections between \SIrange[range-phrase=\mbox{\ and\ }]{1.0e-23}{2.4e-18}{\square\cm}. 
These results are interpreted as constraints on composite dark matter models with two different nucleon-to-nuclear cross section scalings.

\end{abstract}

\maketitle

\section{Introduction}
\label{sec:intro}
Despite the abundance of dark matter (\DM)~\cite{planck_collaboration_planck_2018}, little is known about its particle nature.
While Weakly Interacting Massive Particles (\WIMPs) of electroweak masses and possible thermal origin are promising candidates and are the subject of several recent searches (e.g.~\refscite{deap_collaboration_search_2019,darkside_collaboration_darkside-50_2018,aprile_dark_2018,lux_collaboration_results_2017,pandax-ii_collaboration_dark_2017,supercdms_collaboration_results_2018,PICO}, also Ref.~\cite{APPEC:Billard:2021uyg}), other well-motivated candidates span many orders of magnitude in mass and may evade current constraints.

\DM\ with Planck-scale mass (\mbox{$\WIMPMassSymbol\simeq\SI{e19}{\GeV\per\square\c}$}) may be produced non-thermally, such as in inflaton decay or gravitational mechanisms related to inflation~\cite{kuzmin_matter_1999,kolb_superheavy_2017,chung_superheavy_1998, harigaya_gutzilla_2016,babichev_new_2019}, often related to Grand Unified Theories (GUTs).
Other models describe super-heavy \DM\ produced by primordial black hole radiation~\cite{hooper_dark_2019} or extended thermal production in a dark sector~\cite{kim_super_2019}.

Direct detection constraints at these masses are limited by the \DM\ number density rather than the cross section.  
As a result, even large cross sections permitting multiple scatters remain unconstrained. 
While the finite overburden may allow sufficiently massive particles to be detected underground~\cite{MIMPProposal1:1803.08044}, typical \WIMP\ analyses that reject pileup and multiple-scatter signatures cannot be extrapolated to these high cross sections.
Instead, dedicated analyses are required~\cite{MIMPProposal1:1803.08044,MIMPProposal2:1812.09325,MIMPProposal3:Bramante:2019yss}, which can probe
a variety of theoretical scenarios giving super-heavy, stable, and strongly interacting states~\cite{MIMPProposal2:1812.09325,MIMPModels:ColoredDM,MIMPModels:Nuggets,MIMPModels:BNL:DarkBaryonGeVMediator,MIMPModels:EWSymMonopoles:Bai:2020ttp,MIMPModels:ElectroweakBalls,MIMPModels:BlackHolesSantaCruz}.

Previous direct detection searches constrain \DM\ with \mbox{$\WIMPMassSymbol\lesssim\PreviousDMMassLimit$}~\cite{bernabei_extended_1999,Albuquerque:2003ei,kavanagh_earth-scattering_2018,cappiello_new_2021,Clark:2020mna}.
The present study uses data taken with \DEAP, \SnolabDepth\ underground at \SNOLAB, to probe \WIMPMassSymbol\ up to the Planck scale using multiple-scatter signals, placing the first direct detection constraints at these masses.

\section{Detector, event reconstruction \& data set}
\label{sec:detector}
\DEAP\ contains \larmasserror\ \LAr\ in a spherical acrylic vessel (\AV) with inner surface area \SI{9.1}{\square\meter}, viewed by \AVPMTsNum\ photomultiplier tubes (\PMTs), submerged in a water Cherenkov muon veto (\MV).
Additional details are described in \refscite{amaudruz_design_2019,amaudruz_-situ_2019}. The data acquisition and \WIMP\ search analysis are described in \refscite{deap_collaboration_search_2019,deap-3600_collaboration_first_2018}. 

Energy depositions are measured by counting photoelectrons (\PEs) in the \PMTs\ resulting from \LAr\ scintillation.
\PEs\ are measured by charge-division, as in \refcite{deap-3600_collaboration_first_2018}, rather than the Bayesian algorithm in \refcite{deap_collaboration_search_2019}, as the energies and event topologies of interest extend beyond the latter's validation range.

The pulse shape of a waveform $w(t)$ summed over all \PMTs\ is quantified with \FPrompt, as in \refcite{deap-3600_collaboration_first_2018},
\begin{equation}
    \FPrompt=\frac{\int^{\SI{150}{\ns}}_{\SI{-28}{\ns}}w(t)\,dt}{\int^{\SI{10000}{\ns}}_{\SI{-28}{\ns}}w(t)\,dt}~.
\end{equation}
\FPrompt\ discriminates single-scatter electronic and nuclear recoils~\cite{the_deap_collaboration_pulseshape_2021} and decreases with the number of scatters, separating single- and multiple-scatters with increasing efficiency at high cross sections.

A second discriminator \subeventN\ is calculated with a peak-finding algorithm based on the waveforms' slope and identifies coincident scintillation pulses in a \QPEWindow\ window.
This algorithm best identifies multiple-scatter events when the scatters are spread out in time and produce well-separated peaks.

To reduce the volume of data written to disk due to the \ArThreeNineTotalActivity\ of \isotope{Ar}{39}~\cite{deap_collaboration_electromagnetic_2019,deap_collaboration_search_2019}, a ``pre-scale'' region is defined at low \FPrompt\ for \PrescaleRangeEne\ energies.
Only trigger-level information is recorded for \PrescaleFactor\ of such events, limiting sensitivity to the lowest cross sections of interest in the present analysis.

This search uses a blind analysis of (\MIMPLiveTimeNoUnits)\,live-days of data collected between \MIMPSearchStartDate\ and \MIMPSearchEndDate, excluding \SI{3\pm3}{\micro\second\per trigger} to account for \DM\ signals that may be divided between two recorded traces, a \MIMPOpenDatasetLiveTime\ open physics run, and a \MIMPMuonSidebandLiveTime\ muon-coincidence sideband, composed of events within \MIMPMuonCutWindow\ of \MV\ triggers. 
These open datasets informed the background model and cuts, which were frozen prior to unblinding.

\section{Simulation}
\label{sec:simulation}
\DM\ is simulated via Monte Carlo with the \RAT\ software~\cite{rat}, built upon \Geant~\cite{Geant4}, in two steps: 1) it is attenuated in the overburden, 2) it is propagated in the detector, simulating optical and data acquisition (DAQ) responses.
\DM\ is generated \MIMPAtmRadius\ above the Earth's surface with the Standard Halo Model velocity distribution~\cite{baxter_recommended_2021,lewin_review_1996,smith_rave_2007,mccabe_earthtextquotesingles_2014,schonrich_local_2010,bland-hawthorn_galaxy_2016,abuter_improved_2021} and propagated through the Earth to a \MIMPGenerationRadius\ shell surrounding the \AV.
\DM\ is boosted into the detector's reference frame for a randomized date, following \refscite{kavanagh_earth-scattering_2018,bozorgnia_daily_2011}.

Assuming continuous energy loss, the attenuation of \DM\ at position $\vec{r}$ is calculated numerically as~\cite{MIMPProposal2:1812.09325}
\begin{equation}
    \left\langle \frac{dE_\chi}{dt}\right\rangle(\vec{r}) = -\sum_in_i(\vec{r})\sigma_{i,\chi}\langle E_R\rangle_iv~,
\end{equation}
with $v$ the lab-frame \DM\ speed, 
$n_i$ the number density of nuclide $i$,
$\sigma_{i,\chi}$ the \DM-nucleus scattering cross section,
and $\langle E_R\rangle_i$ the average recoil energy, 
\begin{gather}
\begin{aligned}
    \langle E_R \rangle_i &= \frac{1}{\sigma_{i,\chi}} \int_0^{E_i^\text{max}}E_R\frac{d\sigma_{i,\chi}}{dE_R}dE_R~, \\
    E_i^\text{max} &= [4\WIMPMassSymbol \text{m}_i/(\WIMPMassSymbol+\text{m}_i)^2] E_\chi~,
\end{aligned}
\end{gather}
where 
\WIMPMassSymbol\ and $\text{m}_i$ are the \DM\ and nucleus mass, respectively,  and $d\sigma_{i,\chi}/dE_R$ is the model-dependent differential scattering cross section (see \refsec{sec:interpretations}).

The atmospheric density profile is taken from \refcite{atm}, composed of \mbox{\MIMPAtmNtwoConc\ \ce{N2}} and \mbox{\MIMPAtmOtwoConc\ \ce{O2}}, and the Earth's density profile and composition are from \refscite{Lundberg:2004dn,dziewonski_preliminary_1981}.
Uncertainties in the Earth and atmosphere models negligibly affect the present study.

\DM\ is then propagated through \DEAP.
The detector response is calibrated up to \SI{10}{\MeVee} using \ngamma\ lines from an \ce{^241AmBe} source, giving a factor of \MIMPLYCorrFactor\ used to scale the simulated \PE\ response.

\begin{figure}[t!]
    \centering
    \includegraphics[width=\linewidth]{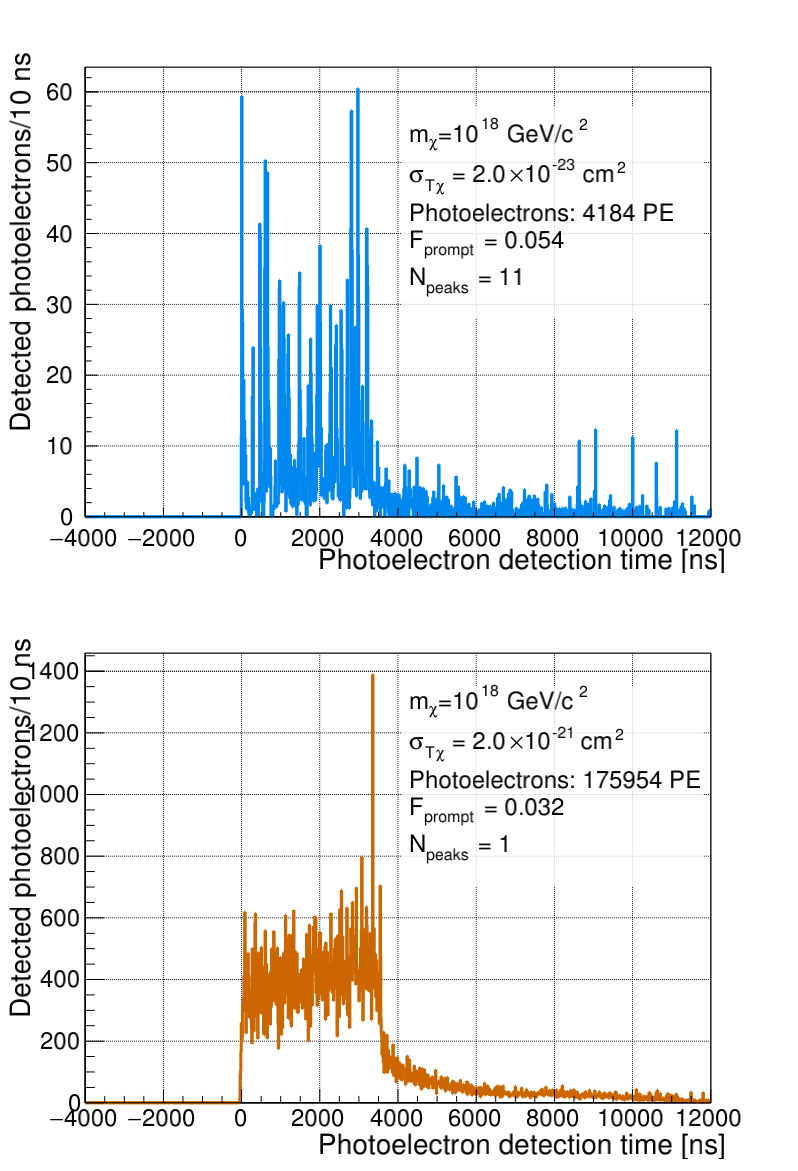}
    \caption{Example simulated \PE\ time distributions for \DM\ with $\WIMPMassSymbol=\SI{e18}{\GeV\per\square\c}$ with low and high $\sigmaTX$. }
    \label{fig:mimpwfs}
\end{figure}

\reffig{fig:mimpwfs} shows two simulated \PE\ time distributions. 
At lower nuclear scattering cross sections (denoted $\sigmaTX$), \subeventN\ counts peaks from individual scatters, which merge at higher $\sigmaTX$, causing it to lose accuracy.
In this regime, the signal energy and duration, typically \SI{<6}{\micro\second}, depend on the \DM\ speed and track length in \LAr,
making \FPrompt\ an estimate of the fraction of scatters in a \TimeFitWindowEnd\ window around the start of the signal, which decreases at higher $\sigmaTX$.

\begin{figure}[t!]
    \centering
    \includegraphics[width=\linewidth]{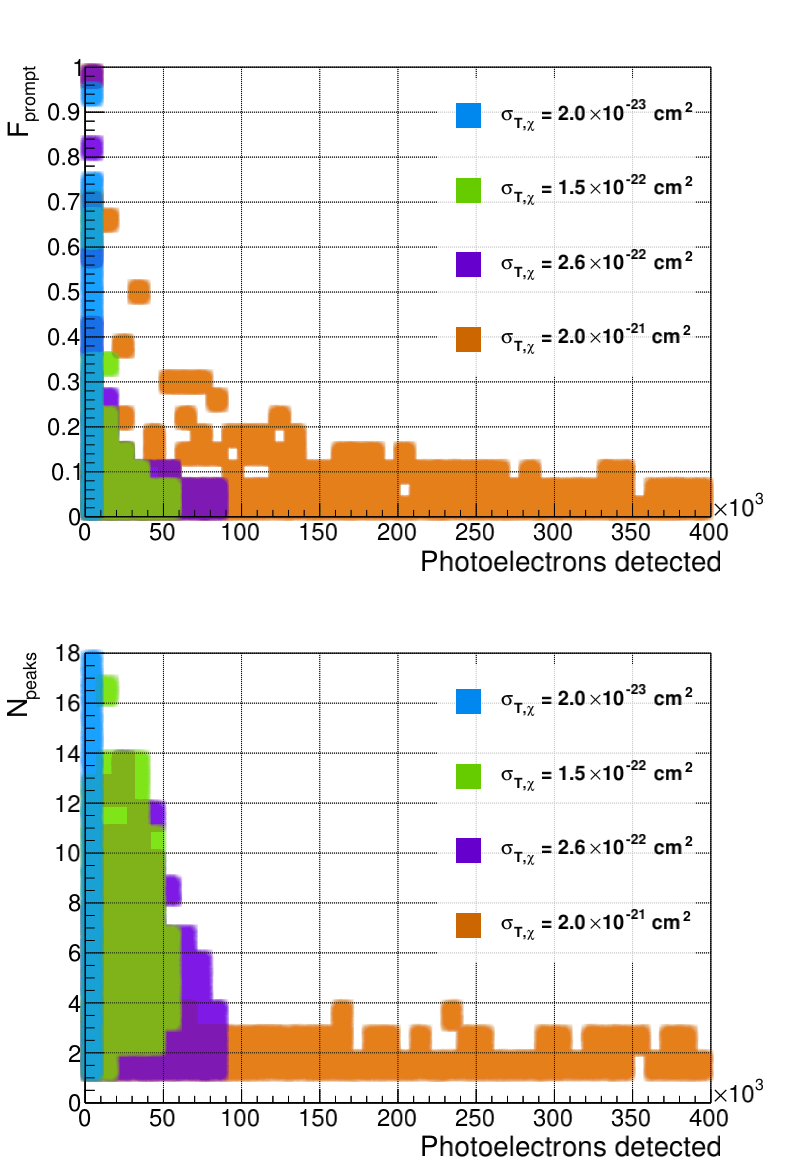}
    \caption{Simulated \FPrompt\ and \subeventN\ distributions for \DM\ with \mbox{\WIMPMassSymbol\SI{=e18}{\GeV\per\square\c}} for various $\sigmaTX$.}
    \label{fig:pileupvars}
\end{figure}

Near $\sigmaTX\simeq\SI{e-23}{\square\cm}$, \subeventN\ grows with increasing $\sigmaTX$ as the \DM\ scatters more times.
As peaks merge, \subeventN\ decreases with $\sigmaTX$, as seen in \reffig{fig:pileupvars}.  
However, \FPrompt\ also decreases and narrows as $\sigmaTX$ grows.
For the simulated $\sigmaTX$, overburden effects have a negligible impact on the \DM\ signal above \SI{e12}{\GeV\per\square\c} and become significant at lower \WIMPMassSymbol.

\section{Analysis and results}
\label{sec:analysis}

\begin{table*}[htb]
\begin{center}
 \begin{tabular}{cllcccc} 
 \hline\hline 
  ROI & \PE\ range &  Energy [\si{\MeVee}]  & $\text{N}_\text{peaks}^\text{min}$ & $\text{F}_\text{prompt}^\text{max}$ & $\NumberExpectedBackgroundSymbol$ & $\text{N}_\text{obs.}$ \T\B \\\hline
 1 & \MIMPROIOnePERange  & \MIMPROIOneEneRange  & \MIMPROIOneSubEventNMin  & \MIMPROIOneFpMax  & \MIMPROIOneNbkgd & \MIMPROIOneNobs \T\\
 2 & \MIMPROITwoPERange  & \MIMPROITwoEneRange  & \MIMPROITwoSubEventNMin  & \MIMPROITwoFpMax  & \MIMPROITwoNbkgd & \MIMPROITwoNobs \\  
 3 & \MIMPROIThreePERange& \MIMPROIThreeEneRange& \MIMPROIThreeSubEventNMin& \MIMPROIThreeFpMax& \MIMPROIThreeNbkgd & \MIMPROIThreeNobs \\
 4 & \MIMPROIFourPERange & \MIMPROIFourEneRange & \MIMPROIFourSubEventNMin & \MIMPROIFourFpMax & \MIMPROIFourNbkgd & \MIMPROIFourNobs \\ 
 \hline\hline
\end{tabular}
\caption{\ROI\ definitions, background expectations \NumberExpectedBackgroundSymbol, and observed event counts $\text{N}_\text{obs.}$ in the \MIMPLiveTimeNoErrors\ exposure. A cut rejecting events in a \MIMPMuonCutWindow\ window surrounding each MV trigger is applied to all \ROIs; low-level cuts requiring that signals be consistent with bulk \LAr\ scintillation are applied to \ROIs~1--3.
The upper energy bound on \ROI~4 is estimated assuming a constant light yield above \SI{10}{\MeVee}, the highest energy at which the detector is calibrated.
}
 \label{tab:roibkgds}
\end{center}
\end{table*}
To identify \DM\ over a wide range of energies and scattering lengths, four regions of interest (\ROIs) are defined with different cuts on \subeventN\ and \FPrompt, summarized in \reftab{tab:roibkgds}.
Cuts for \ROIs~1--3 mitigate pileup backgrounds that are negligible in \ROI~4, which uses minimal cuts that can be evaluated without the full simulation.
Doing this allows for constraints on \DM-nucleon scattering cross sections $\sigmaNX$ that are computationally prohibitive to simulate.

\subsection{Backgrounds and selection cuts}
\label{sec:background}

The primary backgrounds come from uncorrelated pileup of signals produced by radioactivity in detector materials, described in \refcite{deap_collaboration_electromagnetic_2019}.
Correlated backgrounds, such as \isotope{Po}{212} \alpds\ following \isotope{Bi}{212} \betds\ with a \PoTwelveHalfLife\ half-life, are removed by requiring
$\subeventN\,\num{> 2}$ for all energies they may populate.

Pileup was modeled by simulation, validated with a \MIMPAmBeLivetime\ calibration run with an \ce{^241AmBe} source, which emits neutrons at a \AmBeSourceActivity\ rate, and with a \MIMPOpenDatasetLiveTime\ non-blind physics run, testing pileup reconstruction for $\subeventN\,\num{\leq4}$ up to \SI{7.4}{\MeV} and $\subeventN\,\num{\leq5}$ up to \SI{2.6}{\MeV}.
Simulated \subeventN\ distributions agreed to within \MIMPSubeventNMaxDisagreement\ in both datasets.
\ROI~4 relies solely on \FPrompt\ for multi-scatter detection, since \subeventN\ could not be tested at these energies.

Two low-level cuts in \ROIs~1--3 ensure signals are from bulk \LAr\ scintillation: $<$\MIMPCFTTRCut\ of \PE\ must be in \PMTs\ in gaseous Ar, with a \DM\ acceptance of \MIMPAcceptanceCFFTR, and $<$\MIMPFmaxpeCut\ of \PE\ must be in the brightest channel, with a \MIMPAcceptanceFmaxpe\ acceptance.

The dominant backgrounds in \ROIs~1--3 are from pileup. 
Pileup rates decrease with energy, allowing the \subeventN\ threshold to accommodate the decreasing accuracy at higher cross sections.
Pileup is negligible in \ROI~4, where muons produce the dominant backgrounds.
Muons are tagged by the veto.
Untagged muons are rejected by the \FPrompt\ cut, tuned on the muon-coincidence dataset.
The background expectation is determined using the flux in \refcite{Aharmim_2009}.

\begin{figure}[htb]
    \centering
    \includegraphics[width=1.1\linewidth]{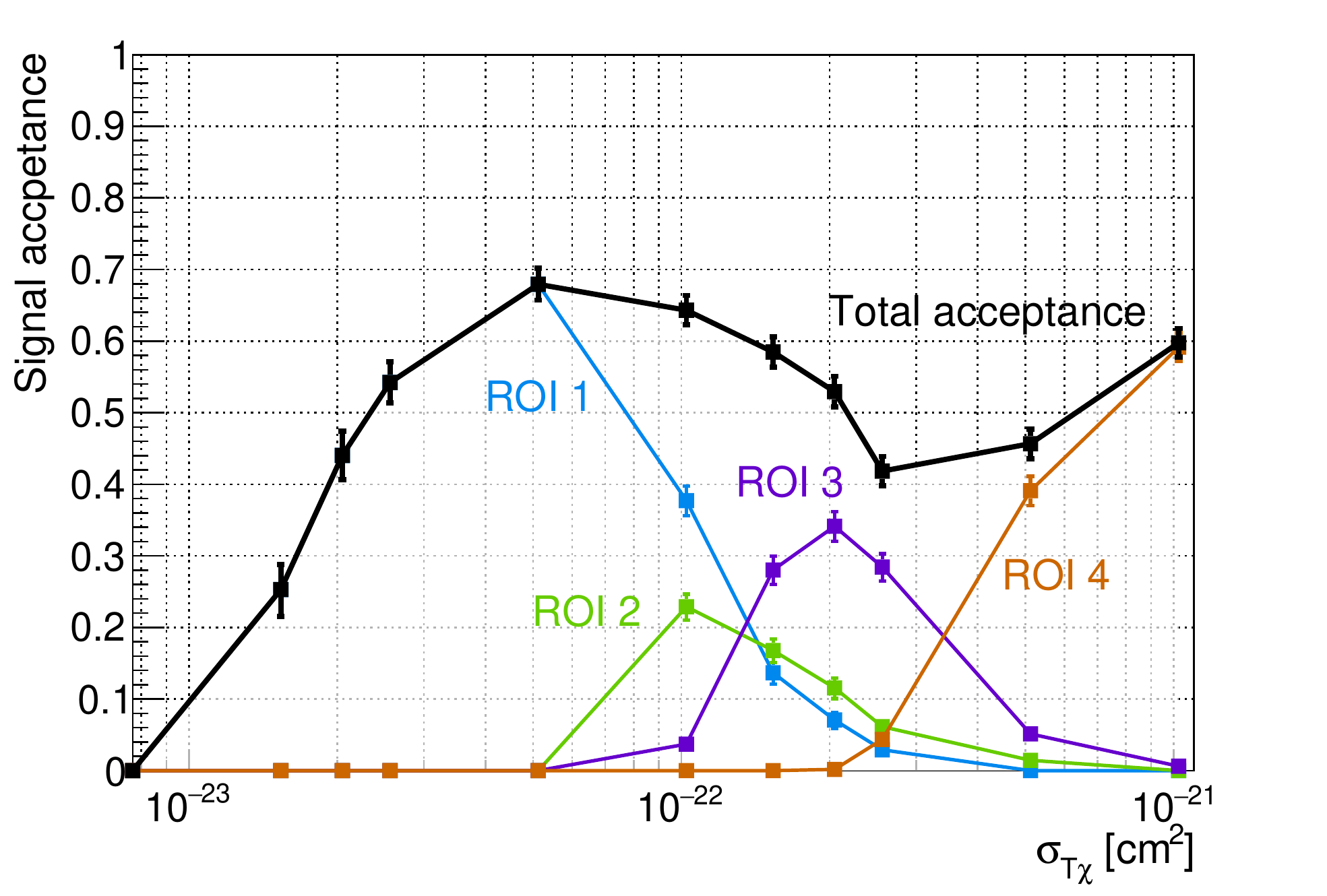}
    \caption{Probability of \DM\ with  \mbox{\WIMPMassSymbol\SI{=e18}{\GeV\per\square\c}} populating each \ROI\ and surviving all cuts at varying $\sigmaTX$.}
    \label{fig:acceptance}
\end{figure}

\reftab{tab:roibkgds} summarizes cuts and backgrounds in each \ROI, defined by the \PE\ range.
Energies are provided for illustrative purposes; the listed upper bound on \ROI~4 assumes the light yield remains constant above \SI{10}{\MeVee}, the maximum energy at which the detector is calibrated.
Its upper \PE\ bound is consistent with the highest scale at which the DAQ system's performance was tested using calibration data collected with a light injection system.
\reffig{fig:acceptance} shows the probability of \SI{e18}{\GeV\per\square\c} \DM\ reconstructing in the \PE\ range for each \ROI\ and passing all cuts. 

\subsection{Results}

After finalizing the selection cuts and background model with a total background expectation of \MIMPTotalBkgdExpectation\ across all \ROIs, the blinded dataset was opened, revealing zero events.
These null results allow any \DM\ model predicting more than \PaperThreeAVSurfaceNEventUpperLimit\ events across all \ROIs\ to be excluded at the \NinetyPerCentCL

The number of events expected in live time $T$ is
\begin{equation}
    \NumberExpectedSignalSymbol = T\int d^3\vec{v} \int dA \frac{\RhoDMSymbol}{\WIMPMassSymbol} |v| f(\vec{v}) \epsilon (\vec{v},\sigmaTX,\WIMPMassSymbol),
\label{eq:nevts}
\end{equation}
with local DM density $\RhoDMSymbol=\RhoDMValue$~\cite{lewin_review_1996}, \DM\ velocity at the detector $\vec{v}$, acceptance $\epsilon$, and surface area $A$. 
\refeqn{eq:nevts} is evaluated by Monte Carlo simulation, including effects detailed in \refsec{sec:simulation}, systematic uncertainties on energy and \subeventN\ reconstruction, and Monte Carlo statistical uncertainties.

\section{Theoretical Interpretations}
\label{sec:interpretations}
The \DM\ signal and $\sigmaNX$-$\sigmaTX$ scaling depend on the \DM\ model.
Two composite models are considered.

\begin{figure}[htb!]
    \centering
    \includegraphics[width=\linewidth]{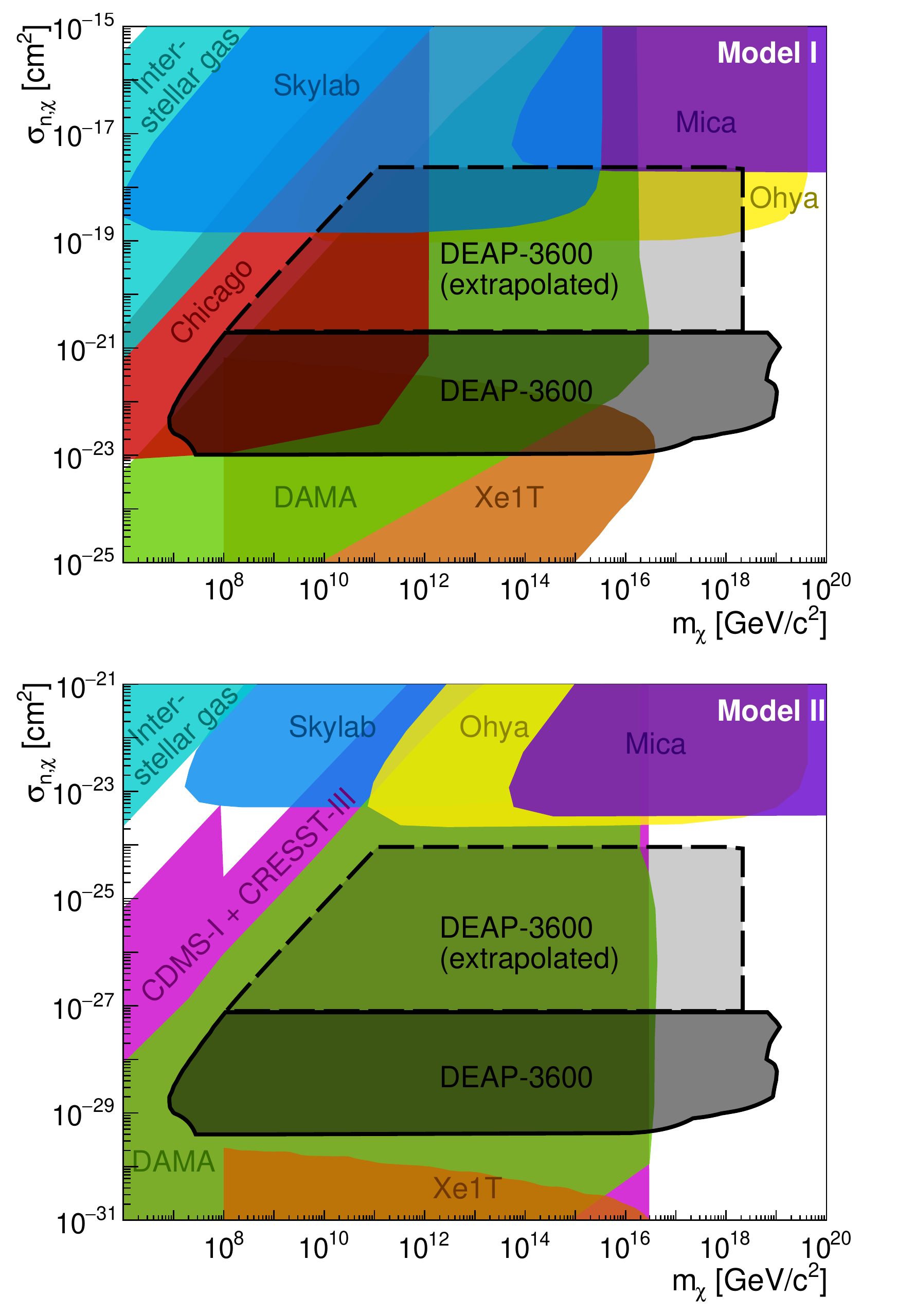}
    \caption{\DM\ masses \WIMPMassSymbol\ and nucleon scattering cross sections $\sigmaNX$ excluded by \DEAP, for Model~I (top) and Model~II (bottom). 
    Extrapolated regions exclude dark matter above the highest simulated cross sections.
    Also shown are other constraints using
    DAMA~\cite{bernabei_extended_1999,BhoonahSkyLab:2020fys},
    interstellar gas clouds~\cite{BhoonahGasClouds:2018gjb,BhoonahGasClouds:2020dzs},
     a recast of CRESST and CDMS-I~\cite{kavanagh_earth-scattering_2018},
    a detector in U.~Chicago~\cite{cappiello_new_2021},
    a XENON1T single-scatter analysis~\cite{Clark:2020mna}, and tracks in the Skylab and Ohya
    plastic etch detectors~\cite{BhoonahSkyLab:2020fys},
    and in ancient mica~\cite{AcevedoMica:2021tbl}.
    Limits from MAJORANA DEMONSTRATOR~\cite{Clark:2020mna} are not shown as the corresponding regions are already excluded by DAMA and XENON1T.
    }  \label{fig:exclusion_I}
    \label{fig:exclusion_II}
    \label{fig:exclusion_III}
\end{figure}

For each model, \NumberExpectedSignalSymbol\ is determined at several \WIMPMassSymbol\ and $\sigmaNX$, and
exclusion regions are built accounting for uncertainties as prescribed in \refcite{cousins_incorporating_1992}.
Upper bounds on \WIMPMassSymbol\ are interpolated with a $\RhoDMSymbol/\WIMPMassSymbol$ flux scaling;
lower bounds are set to the value at which the overburden calculation predicts that \SI{90}{\percent} of expected \DM\ signals will be below \SI{1}{\MeVee} after quenching.
Upper bounds on $\sigmaNX$ are set by the lowest simulated values that that can be excluded, while lower bounds are limited by the highest $\sigmaNX$ that were computationally possible to simulate, $\sigmaNX^\text{max}$. 
At higher $\sigmaNX$, the continuous scattering approximation and the time-of-flight in \LAr\ imply a lower bound on the \ROI~4 acceptance of \MIMPMinimumFpAcceptanceROIFour.
Conservatively treating the probability of reconstructing in \ROI~4 as constant above $\sigmaNX^\text{max}$ and scaling the flux as $\RhoDMSymbol/\WIMPMassSymbol$, exclusion regions are extrapolated to \WIMPMassSymbol\ consistent with null results.
Upper bounds on $\sigmaNX$ are set to $\sigmaNX^\text{max}\times\left(\PE_\text{max}^\text{ROI4}/\PE^\text{sim}_{90}\right)$, where $\PE_\text{max}^\text{ROI4}$ is the upper \PE\ bound of \ROI~4 and $\PE^\text{sim}_{90}$ is the \SI{90}{\percent} upper quantile on the \PE\ distribution at $\sigmaNX^\text{max}$.
These constraints are labeled  ``extrapolated'' in \reffig{fig:exclusion_I}.

\subsection{Model I}

In this model, \DM\ is opaque to the nucleus, so that 
the scattering cross section at zero momentum transfer $q$ is the geometric size of the \DM\ regardless of the target nucleus.
More generally,
\beq
  \frac{d\sigmaTX}{d E_R} = \frac{d \sigmaNX}{d E_R}|F_{\rm T}(q)|^2~,
  \label{eq:xsnuclearcompositeI}
\eeq
where $F_{\rm T}(q)$ is the Helm form factor~\cite{Helm:PhysRev.104.1466,helmFF:Vietze:2014vsa}.
This scaling may give conservative limits for strongly interacting composite \DM~\cite{DigmanBeacomPeterBarn}.
The region excluded for this model is shown in \reffig{fig:exclusion_I} (top).
Here (and in the bottom panel) the lower and upper boundaries are flat because, unlike in \WIMP\ searches where these exclusion $\sigmaTX \propto \mdark$ at high \DM\ masses, the cross section sensitivity is only dependent on the detector's multi-scatter acceptance. 
The right-hand boundary is nearly vertical due to the drop in \DM\ flux with increasing $\mdark$; above the notch is the region where the Earth overburden is dominated by the crust. 
On the left-hand boundary $\sigmaTX \propto \mdark$ due to attenuation in the overburden.

\subsection{Model II}

In this scenario the cross section scales as 
\begin{gather}
\begin{aligned}
    \frac{d\sigmaTX}{d E_R}&= \frac{d\sigmaNX}{d E_R}\bigg(\frac{\mu_{T\chi}}{\mu_{n\chi}}\bigg)^2A^2|F_{\rm T}(q)|^2 \\
    &\simeq \frac{d\sigmaNX}{d E_R} A^4|F_{\rm T}(q)|^2,
  \label{eq:xsnuclearcompositeII}
\end{aligned}
\end{gather}
where $\mu_{\{n,T\}\chi}$ is the \{nucleon, target\}-\DM\ reduced mass and
$A$ is the target mass number.
The excluded region is shown in \reffig{fig:exclusion_II} (bottom).

\refeqn{eq:xsnuclearcompositeII} is the most commonly used scaling, allowing for comparisons with other experiments and with single-scatter constraints.
It may arise from nuclear \DM\ models, outlined in \refscite{compsite_hardylasenby,composite_monroe},
which describe a dark nucleus with $\Ndark$ nucleons of mass $\mdark$ and radius $\rdark$, yielding a total mass \mbox{$\WIMPMassSymbol=\Ndark \mdark$} and radius
$\Rdark=\Ndark^{1/3} \rdark~.$
For $\WIMPMassSymbol \gg \mT$, 
\beq
  \frac{d \sigmaTX}{d E_R}  = \frac{d\sigma_{\rm nD}}{d E_R}\Ndark^2 |F_{\chi}(q) |^2 A^4 |F_{\rm T}(q) |^2~,
  \label{eq:xsnuclearcompositeIII}
\eeq
where $\sigma_{\rm nD}$ is the nucleon-dark nucleon scattering cross section.
To preserve the Born approximation, \refeqn{eq:xsnuclearcompositeIII} is bounded by the geometric cross section:
\beq
\sigmaTX \leq \sigma_{\rm geo} (= 4\pi \Rdark^2 = 4\pi \Ndark^{2/3} \rdark^2)~.
\label{eq:siggeo}
\eeq
For dark nuclei of size $\Rdark \gg \SI{1}{\femto\meter}$, we may identify $\sigmaNX = \Ndark^2 \sigma_{\rm nD}$ for potentials that give rise to $|F_{\chi}(q)|^2 \simeq 1$, and \reffig{fig:exclusion_II} could then constrain such nuclear \DM\ in regions satisfying \refeqn{eq:siggeo}.
We leave detailed studies of such possibilities to future work.

\section{Summary and scope}

This study uses \DEAP\ data to derive new constraints on composite \DM, including the first direct detection results probing Planck-scale masses.
These constraints were obtained through a dedicated analysis of multiple-scatter signals, accounting for the attenuation that the \DM\ would experience in the laboratory's overburden.
The analysis used to achieve these results represents the first study of this kind in a tonne-scale direct detection experiment, extending Planck-scale limits from ancient mica~\cite{AcevedoMica:2021tbl} and etched plastic studies~\cite{BhoonahSkyLab:2020fys} to lower cross sections.

The high-mass sensitivity achieved by \DEAP\ was possible due to its large cross sectional area, which provides a large net to catch dilute \DM.
As a result, limits were placed on two classes of \DM\ models describing  strongly interacting, opaque composites and dark nuclei motivated by the QCD scale with a spherical top-hat potential.

This analysis may be extended to superheavy \DM\ depositing energy via modes other than elastic scattering, (\eg~\refcite{MIMPModels:BlackHolesSantaCruz}), 
to future \LAr, liquid xenon, and bubble chamber detectors,
and to large-scale liquid scintillator (\eg~\SNOp, \JUNO)~\cite{MIMPProposal2:1812.09325} and segmented detectors (\eg~\MATHUSLA)~\cite{MIMPProposal3:Bramante:2019yss}.

\label{sec:conclusions}

\section*{Acknowledgments}

We gratefully acknowledge fruitful interactions with 
Javier Acevedo,
Joe Bramante, and
Alan Goodman.
TRIUMF receives federal funding via a contribution agreement with the National Research Council Canada.
We thank the Natural Sciences and Engineering Research Council of Canada,
the Canadian Foundation for Innovation (CFI),
the Ontario Ministry of Research and Innovation (MRI), 
and Alberta Advanced Education and Technology (ASRIP),
Queen's University,
the University of Alberta,
Carleton University,
the Canada First Research Excellence Fund,
the Arthur B.~McDonald Canadian Astroparticle Research Institute,
DGAPA-UNAM (PAPIIT No.~IN108020) and Consejo Nacional de Ciencia y Tecnolog\'ia (CONACyT, Mexico, Grant A1-S-8960),
the European Research Council Project (ERC StG 279980),
the UK Science and Technology Facilities Council (STFC) (ST/K002570/1 and ST/R002908/1),
the Leverhulme Trust (ECF-20130496),
the Russian Science Foundation (Grant No. 21-72-10065),
the Spanish Ministry of Science and Innovation (PID2019-109374GB-I00), 
and the International Research Agenda Programme AstroCeNT (MAB/2018/7)
funded by the Foundation for Polish Science (FNP) from the European Regional Development Fund.
Studentship support from
the Rutherford Appleton Laboratory Particle Physics Division,
STFC and SEPNet PhD is acknowledged.
We thank SNOLAB and its staff for support through underground space, logistical, and technical services.
SNOLAB operations are supported by the CFI
and Province of Ontario MRI,
with underground access provided by Vale at the Creighton mine site.
We thank Vale for their continuing support, including the work of shipping the acrylic vessel underground.
We gratefully acknowledge the support of Compute Canada,
Calcul Qu\'ebec,
the Centre for Advanced Computing at Queen's University,
and the Computation Centre for Particle and Astrophysics (C2PAP) at the Leibniz Supercomputer Centre (LRZ)
for providing the computing resources required to undertake this work.

\bibliographystyle{deap}
\bibliography{refs}

\end{document}